# Machine Ethics and Automated Vehicles

**Author: Noah J. Goodall**

**Abstract**

Road vehicle travel at a reasonable speed involves some risk, even when using computer-controlled driving with failure-free hardware and perfect sensing. A fully-automated vehicle must continuously decide how to allocate this risk without a human driver's oversight. These are ethical decisions, particularly in instances where an automated vehicle cannot avoid crashing. In this chapter, I introduce the concept of moral behavior for an automated vehicle, argue the need for research in this area through responses to anticipated critiques, and discuss relevant applications from machine ethics and moral modeling research.

## 1 Ethical Decision Making for Automated Vehicles

Vehicle automation has progressed rapidly this millennium, mirroring improvements in machine learning, sensing, and processing. Media coverage often focuses on the anticipated safety benefits from automation, as computers are expected to be more attentive, precise, and predictable than human drivers. Mentioned less often are the novel problems from automated vehicle crash. The first problem is liability, as it is currently unclear who would be at fault if a vehicle crashed while self-driving. The second problem is the ability of an automated vehicle to make ethically-complex decisions when driving, particularly prior to a crash. This chapter focuses on the second problem, and the application of machine ethics to vehicle automation.

Driving at any significant speed can never be completely safe. A loaded tractor trailer at 100 km/hr requires eight seconds to come to a complete stop, and a passenger car requires three seconds [1]. Truly safe travel requires accurate predictions of other vehicle behavior over this time frame, something that is simply not possible given the close proximities of road vehicles.

To ensure its own safety, an automated vehicle must continually assess risk: the risk of traveling a certain speed on a certain curve, of crossing the centerline to pass a cyclist, of side-swiping an adjacent vehicle to avoid a runaway truck closing in from behind. The vehicle (or the programmer in advance) must decide how much risk to accept for itself and for the adjacent vehicles. If the risk is deemed acceptable, it must decide how to apportion this risk among affected parties. These are ethical questions that, due to time constraints during a crash, must be decided by the vehicle autonomously.

The remainder of the chapter is organized into the parts. In section 2, responses are provided to nine criticisms of the need for ethics research in automated vehicle decision systems. Section 3 contains reviews of relevant ethical theories and moral modeling research. The chapter is summarized in Section 4.

## 2 Criticisms of the Need for Automated Vehicle Ethics Systems, and Responses

Future automated vehicles will encounter situations where the "right" action is morally or legally ambiguous. In these situations, vehicles need a method to determine an ethical action. However, there is disagreement among experts on both of these points. This section lists nine criticisms of the importance of ethics in vehicle automation, with responses to each.

**Criticism 1: Automated vehicles will never (or rarely) crash.** If an automated vehicle never crashes, then there is no need to assess or assign risk because driving no longer contains risk. Industry experts are mostly cautious regarding whether vehicle automation can ever eliminate all crashes. Claims of complete safety are often based on assumptions about the capabilities automated vehicles and their environments. These assumptions can be grouped into three scenarios: automated vehicles with imperfect systems, automated vehicles with perfect systems driving in mixed traffic with human drivers, and automated vehicles with perfect systems driving exclusively with other automated vehicles. Crashes are possible in each scenario, as described in the following paragraphs.

▶    **Imperfect systems.** Any system ever engineered has occasionally failed. In the realm of automated vehicle, Fraichard and Kuffner list four reasons for a collision: hardware failures, software bugs, perceptual errors, and reasoning errors [2]. While hardware failures may be somewhat predictable and often gradual, software failures are unexpected and sudden, and may prove riskier at high speeds. Perceptual errors may result in misclassifying an object on the roadway. Even if a pedestrian is correctly classified, an automated vehicle would need some way to perceive her intent, e.g. whether she is about to step into the road or is merely standing on the sidewalk. A mistake in this calculation could lead to a crash, especially considering the close proximity and high speed differentials on roadways.

▶ **Perfect systems with mixed human-driven traffic.** A perfect automated vehicle with complete awareness of its surroundings should be able to safely avoid static objects. Dynamic objects with unpredictable behavior pose a greater challenge. The best way to avoid a collision is to avoid any place, time, and trajectory on the roadway (referred to as a state) which could possibly lead to a crash. In robotics, a state where all possible movement result in a crash is referred to as an inevitable collision state [3]. Researchers have acknowledged that with road vehicles, there is no way to completely avoid inevitable collision states [4], only to minimize the probability of entering one [5]. The only reasonable strategy is to construct a model of the expected behavior of nearby vehicles and try to avoid likely collisions—based on patent filings, this appears to be a component of Google's self-driving car [6]. Without a sophisticated model of expected vehicle behavior, a "safe" automated vehicle would be forced to overreact to perceived threats. For example, a "flying pass" maneuver, where a vehicle approaches a stopped queue at high speed only to move into a dedicated turn lane at the last moment, appears identical to a pre-crash rear-end collision [7, p. 140]. To guarantee safety, an automated vehicle would have to evade many similar maneuvers each day. This is both impractical and dangerous.

▶ **Perfect systems without human-driven traffic.** Perfect vehicles traveling on a freeway with other perfect vehicles should be able to safely predict each other's behavior and even communicate wirelessly to avoid collisions. Yet these vehicles would still face threats from wildlife (256,000 crashes in the U.S. in 2000), pedestrians (73,000 crashes), and bicyclists (51,000 crashes) [8]. Although a sophisticated automated vehicle would be safer than a human driver, some crashes may be unavoidable. Furthermore, the perfect systems described in this scenario are neither likely nor near-term.

**Criticism 2: Crashes requiring complex ethical decisions are extremely unlikely.** In order to demonstrate the difficulty of some ethical decisions, philosophers will use examples that seem unrealistic. The trolley problem [9], where a person must decide whether to switch the path of a trolley onto a track that will kill one person in order to spare five passengers, is a common example [10]. The trolley problem is popular because it is both a difficult problem and one where people's reactions are sensitive to context, e.g. pushing a person onto the track instead of throwing a switch produces different responses, even though the overall outcome is the same.

The use of hypothetical examples may suggest that ethics are only needed in incredibly rare circumstances. However a recent profile of Google's self-driving car team suggests that ethics are already being considered in debris avoidance: "What if a cat runs into the road? A deer? A child? There were moral questions as well as mechanical ones, and engineers had never had to answer them before" [11]. Ethical decisions are needed whenever there is risk, and risk is always present when driving.

One can argue that these are simple problems, e.g. avoid the child at all costs and avoid the cat if safe to do so. By comparison, however, the trolley problem is actually fairly straight-forward—it has only one decision, with known consequences for each alternative. This is highly unrealistic. A vehicle faces decisions with unknown consequences, uncertain probabilities of future actions, even uncertainty of its own environment. With these uncertainties, common ethical problems become "complex" very quickly.

**Criticism 3: Automated vehicles will never (or rarely) be responsible for a crash.** This assumes that absence of liability is equivalent to ethical behavior. Regardless of fault, an automated vehicle should behave ethically to protect not only its own occupants, but also those at fault.

**Criticism 4: Automated vehicles will never collide with another automated vehicle.** This assumes that an automated vehicle's only interactions will be with other automated vehicles. This is unlikely to happen in the near future for two reasons. First, the vehicle fleet is slow to turn over. Even if every new vehicle sold in the U.S. was fully-automated, it would be 30 years before 90% percent of vehicles were replaced [12]. Second, unless automated vehicle-only zones are established, any fully-automated vehicle will have to interact with human drivers, pedestrians, bicyclists, motorcyclists, and trains. Even an automated-only zone would encounter debris, wildlife, and inclement weather. These are all in addition to a vehicle's own hardware, software, perceptual, and reasoning failures. Any of these factors can contribute to or independently cause a crash.

**Criticism 5: In level 2 and 3 vehicles, a human will always be available to take control, and therefore the human driver will be responsible for ethical decision making.** Although the National Highway Traffic and Safety Administration (NHTSA) definitions require that a person be available to take control of a vehicle with no notice in a level 2 automated vehicle and within a reasonable amount of time in a level 3 automated vehicle [13], this may be an unrealistic expectation for most drivers.

In a level 2 vehicle, this would require that a driver pay constant attention to the roadway, similar to when using cruise control. Drivers in semi-autonomous vehicles with lane-keeping abilities on an empty test track exhibited significant increases in eccentric head turns and secondary tasks during automated driving, even in the presence of a researcher [14]. Twenty-five percent of test subjects were observed reading while the vehicle was in autonomous mode. Similar results have been found in driving simulator studies [15]. The effect of automation on a driver's attention level remains an open question, but early research suggests that a driver cannot immediately take over control of the vehicle safely. Most drivers will require some type of warning time.

Level 3 vehicles provide this warning time, but the precise amount of time needed is unknown. The NHTSA guidance does not specify an appropriate warning time [13], although some guidance can be found in highway design standards. The American Association of State Highway and Transportation Officials (AASHTO) recommends highway designers allow 200 to 400 meters for a driver to perceive and react to an unusual situation at 100 km/hr [16]. This corresponds to 7 to 14 seconds, much of which is beyond the range of today's radar at 9 seconds [17]. In an emergency, a driver may be unable to assess the situation and make an ethical decision within the available time frame. In these situations, the automated vehicle would maintain control of the vehicle, and by default be responsible for ethical decision making.

**Criticism 6: Humans rarely make ethical decisions when driving or in crashes, and automated vehicles should not be held to the same standard.** Drivers may not believe themselves to be making ethical decisions while driving, but they actually make these decisions often. The decision to speed or to cross a yellow line to provide a cyclist additional room are examples of ethical decisions. Any activity that transfers risk from one person to another involves ethics, and automated vehicles should be able to make acceptable decisions in similar environments. Considering that Americans drive 4.8 trillion kilometers each year [18], novel situations requiring ethics should emerge steadily.

**Criticism 7: An automated vehicle can be programmed to follow the law, which will cover ethical situations.** Existing laws are not nearly comprehensive or specific enough to produce reasonable actions in a computer. Lin provides an example of an automated vehicle coming across a tree branch in the road. If there was no oncoming traffic, a reasonable person would cross the double yellow line to get around the tree, but an automated vehicle programmed to follow the law would be forced to wait until the branch was cleared [19].

Of course, laws could be added for these types of situations. This can quickly become a massive undertaking—one would need computer-understandable definitions of terms like "obstruction" and "safe" for an automated vehicle whose perception system is never completely certain of anything. If enough laws were written to cover the vast majority of ethical situations, and they were written in such a way as to be understood by computers, then the automated vehicle ethics problem would be solved. Current law is not close to these standards.

**Criticism 8: An automated vehicle should simply try to minimize damage at all times.** This proposes a utilitarian ethics system, which is addressed in section 3.1 and in previous work [20]. Briefly, utilitarianism's main obstacle is that it does not recognize the rights of individuals. An utilitarian automated vehicle given the choice between colliding with two different vehicles would select the one with the higher safety rating. Although this would maximize overall safety, most would consider it unfair.

**Criticism 9: Overall benefits outweigh any risks from an unethical vehicle.** This is perhaps the strongest argument against automated vehicle ethics research, that any effort which may impede the progress of automation indirectly harms those who die in the interim between immediate and actual deployment.

While preliminary evidence does not prove automation is safer than human drivers [20], it seems likely that automation will eventually reduce the crash rate. Lin has argued, however, that a reduction in overall fatalities may be considered unethical [21], as improved safety for one group may come at the expense of another. If vehicle fatalities are reduced, but cyclist fatalities increase, even an overall safety improvement might be unacceptable to society.

Second, this assumption uses a purely utilitarian view that maximizing lives saved is the preferred option. Society, however, often uses a different value system considering the context of a given situation. For example, the risk of death from nuclear meltdown is often over-valued, while traffic fatalities are undervalued. Society may disagree that a net gain in safety is worth a particularly frightening risk. If, in fact, the ultimate goal is to improve safety, then ensuring that automated vehicles behave in acceptable ways is critical to earning the public's trust of these new technologies.

Finally, the safety benefits of automated vehicles are still speculative. To be considered safer than a human driver with 99% confidence, an automated passenger vehicle would need to travel 1.1 million kilometers without crashing and 482 million kilometers without a fatal crash [20]. As of this writing, an automated vehicle has yet to safely reach these mileages.

## 3    Relevant Work in Machine Ethics and Moral Modeling

There are two main challenges when formulating an ethical response for an automated vehicle. The first is to articulate society's values across a range of scenarios. This is especially difficult given that most research into morality focuses on single choices with known outcomes (one person will always die if the trolley changes track), while in reality outcomes are uncertain and there are several layers of choices. The second challenge is to translate these morals into language that a computer can understand without a human's ability to discern and analogize.

The recent field of machine ethics addresses these challenges through the development of artificial autonomous agents which can behave morally. While much of machine ethics work is theoretical, a few practical applications include computer modeling of human ethics in areas such as medicine, defense, and engineering. This section provides background on ethical theories, and reviews examples of computational moral modeling.

**3.1     Ethical Theories**

Researchers have investigated the potential for various moral theories for use in machine ethics applications, including utilitarianism [22], Kantianism [23]–[25], Smithianism [26], and deontologicalism [27], [28]. Deontologicalism and utilitarianism have been discussed as potentials for automated vehicle ethics, with shortcomings found with both theories [20].

Deontological ethics consist of limits that are placed on a machine's behavior, or a set of rules that it cannot violate. Asimov's three laws of robotics are a well-known example of deontological ethics [29]. A shortcoming of deontological ethics appears when reducing complex human values into computer code. Similar to the traffic law example from this chapter's seventh criticism, rules generally require some common sense in their application, yet computers are only capable of literal interpretations. These misinterpretations can lead to unexpected behavior. In Asimov's laws, an automated vehicle might avoid braking before a collision because this action would first give its occupants whiplash, thereby violating the first law prohibiting harm to humans. Rules can be added or clarified to cover different situations, but it is unclear if any set of rules could encompass all situations. Developing rules also requires that someone articulate human morals, an exceptionally difficult task given that there has never been complete agreement on the question of what is right and wrong.

Another useful moral theory is utilitarianism. This dictates that an action is moral if the outcome of that an action—or in the case of automated vehicles, the expected outcome—maximizes some utility. The advantage of this method is that it is easily computable. However, it is difficult to define a metric for the outcome. Property damage estimates can produce unfair outcomes, as they would recommend colliding with a helmeted motorcyclist over a non-helmeted one, as the helmeted rider is less likely experience costly brain damage. This example illustrates another shortcoming of utilitarianism—it generally maximizes the collective benefit rather than individuals' benefits, and does not consider equity. One group may consistently benefit (un-helmeted riders) while another loses.

Hansson has noted that risk-taking in radiation exposure combines the three main ethical theories of virtue (referred to as justification), utilitarianism (optimization), and deontologicalism (individual dose limits) [30]. Automated vehicle ethics will also likely require a combination of two or more ethical theories.

**3.2     Practical Applications**

There have been several attempts to develop software that can provide guidance in situations requiring ethics. One of the first examples was a utilitarian software tool called Jeremy [31]. This program measured the utility of any action's outcome by using the straightforward product of the outcome's utility intensity, duration, and probability, each of which were estimated by the user. In an automated vehicle environment, utility could be defined as safety or the inverse of damage costs, with intensity, duration, and probability estimated from crash models. A major shortcoming of this model is its exclusive use of utilitarianism, an ethical theory which disregards context, virtues, and limits on individual harm.

The team behind Jeremy later introduced two other software tools. The first was W.D. [31], which used a duty-based ethical theory influenced by William D. Ross [32] and John Rawls [33]. This was followed by a similar program MedEthEx [34], a tool meant for medical applications and reflecting the duties identified in Principles of Biomedical Ethics [35]. Both of these program are deontological, and are trained using test cases that either violate or adhere to a formulated set of duties as indicated an integer score. The software uses machine learning to determine whether test cases of action are moral or immoral based on adherence to ethical principles, and calibrates these assessments using expert judgment. The output provides an absolute conclusion whether an action is right or wrong, and indicates which ethical principles were most important in the decision.

McLaren has developed two tools to aid in ethical decision making. The first tool is Truth-Teller, a program that analyzes two case studies where the subject must decide whether or not to tell the truth [36]. The program identifies similarities and differences between the cases, and lists reasons for or against telling the truth in each situation. This is an example of casuistic reasoning, where one reaches a conclusion by comparing a problem with similar situations instead of using rules learned from a set of test cases. Case studies are input using symbols rather than natural language processing to be more easily machine-readable. A similar program from McLaren, SIROCCO [36], uses casuistry to identify principles from the National Society of Professional Engineers code of ethics relevant to an engineering ethics problem. Like Truth-Teller, SIROCCO avoids moral judgments, and instead suggests ethically relevant information that can help a user make decisions.

The U.S. Army recently funded research into automated ethical decision making as a support tool for commanders and eventual use in robotic systems. The first step in this effort is a computer model which attempts to assess the relative morality of two competing actions in a battlefield environment. This model, referred to by its developers as the Metric of Evil, attempts to "provide results that resemble human reasoning about morality and evil" rather than replicate the process of human reasoning [37]. To calculate the Metric of Evil, the model sums the evil for each individual consequence of an action, taking into account high and low estimates of evil, confidence intervals, and intentionality. A panel of experts then rates a set of ethical test cases, and the weights of each type of consequence are adjusted so that the model output matches expert judgment. While the Metric of Evil provide decisions on which action is more ethical, it does not provide the user with evidence supporting its conclusion.

Computational moral modeling is in its infancy. The efforts described in this chapter, particularly MedEthEx and the Metric of Evil, show that it is possible to solve ethical problems automatically, although much work is needed, particularly in model calibration and incorporating uncertainty.

## 4 Summary

Automated vehicles, even sophisticated examples, will continue to crash. To minimize damage, the vehicle must continually assess risk to itself and others. Even simple maneuvers will require the vehicle to determine if the risk to itself and other is acceptable. These calculations, the acceptance and apportionment of risk, are ethical decisions, and human drivers will not be able to oversee these decisions. The vehicle must at times make ethical choices autonomously, either via explicit pre-programmed instructions, a machine learning approach, or some combination of the two. The fields of moral modeling and machine ethics has made some progress, but much work remains. This chapter is meant as a guide for those first encountering ethical systems as applied in automated vehicles to help frame the problem, convey core concepts, and provide directions for useful research in related fields.

**Full Authors' Information**

Noah J. Goodall, Ph.D., P.E.
Virginia Center for Transportation Innovation and Research
530 Edgemont Road
Charlottesville, VA

United States
E-mail: noah.goodall@vdot.virginia.gov